\documentclass[letterpaper, 10 pt, conference]{ieeeconf}  

\IEEEoverridecommandlockouts                              
\usepackage[T1]{fontenc}
\usepackage{fontawesome}
\usepackage{amsmath, amssymb}
\usepackage{amsthm}
\usepackage{import}
\usepackage[capitalize, nameinlink]{cleveref}
\crefformat{equation}{(#2#1#3)}
\crefrangeformat{equation}{#3(#1)#4--#5(#2)#6}
\crefmultiformat{equation}{(#2#1#3)}{ and~(#2#1#3)}{, (#2#1#3)}{ and~(#2#1#3)}
\usepackage{eucal}
\usepackage{bm}
\usepackage{mathtools}
\usepackage{braket}
\usepackage{physics}
\usepackage{algpseudocode, algorithm}

\usepackage{enumerate}
\usepackage{adjustbox}
\usepackage{xcolor}
\colorlet{secret}{black!20}
\usepackage{tikz}
\usetikzlibrary{automata, positioning, arrows, arrows.meta}
\tikzset{
  ->,
  >=Latex,
  auto,
  node distance=20mm,
  on grid,
  inner sep=2pt,
  bend angle=20,
  every state/.style={thick, minimum size=6mm},
  initial text=,
}
\usepackage{xkvltxp}
\usepackage[nomargin, inline, index]{fixme}

\DeclareMathOperator{\disjoint}{\dot\cup}
\DeclareMathOperator*{\bigdisjoint}{\dot\bigcup}
\DeclareMathOperator{\supc}{\sup\mathcal{C}}
\DeclareMathOperator{\power}{Pwr}
\DeclareMathOperator{\sand}{\&}

\theoremstyle{definition}
\newtheorem{thm}{Theorem}
\crefname{thm}{Theorem}{Theorems}
\newtheorem{lem}[thm]{Lemma}
\crefname{lem}{Lemma}{Lemmas}
\newtheorem{defn}[thm]{Definition}
\crefname{defn}{Definition}{Definitions}

\newtheorem{prop}[thm]{Proposition}
\crefname{prop}{Proposition}{Propositions}
\newtheorem{exmp}[thm]{Example}
\crefname{exmp}{Example}{Examples}

\newtheorem{prob}[thm]{Problem}
\crefname{prob}{Problem}{Problems}

\crefname{rem}{Remark}{Remarks}

\title{\LARGE \textbf{Secret Securing with Multiple Protections and Minimum
    Costs}%
}

\author{Shoma Matsui and Kai Cai
  \thanks{S. Matsui and K. Cai are with the Department of Electrical and
    Information Engineering at Osaka City University, Osaka, Japan.
    S. Matsui: \texttt{matsui\_s@c.info.eng.osaka-cu.ac.jp}, K. Cai:
    \texttt{kai.cai@eng.osaka-cu.ac.jp}}%
}%

\begin{document}

\listoffixmes

\maketitle
\thispagestyle{empty}
\pagestyle{empty}

\begin{abstract}

  In this paper we study a security problem of protecting secrets with multiple
  protections and minimum costs. The target system is modeled as a
  discrete-event system (DES) in which a few states are secrets, and there are
  multiple subsets of protectable events with different cost levels. We
  formulate the problem as to ensure that every string that reaches a secret
  state (from the initial state) contains a specified number of protectable
  events and the highest cost level of these events is minimum. We first provide
  a necessary and sufficient condition under which this security problem is
  solvable, and then propose an algorithm to solve the problem based on the
  supervisory control theory of DES. The resulting solution is a protection
  policy which specifies at each state which events to protect and the highest
  cost level of protecting these events is minimum. Finally, we demonstrate the
  effectiveness of our solution with a network security example.

\end{abstract}

\section{Introduction}\label{sec:introduction}

Various security problems called cybersecurity issues have attracted much
interest of researchers. In real systems, security problems can be caused by
administrators' mistakes or vulnerabilities of products.
\cite{brooks2018cybersecurity}~introduces practical and technical methods
relevant to security issues in the real world. In general, there are some
secrets in the system which intruders want to steal without raising an alert,
namely without being detected, and such secrets must be protected against
malicious access of intruders. At the same time, the cost to protect secrets
must be taken into account because infinite protection cost is infeasible in
practice.

In this paper, we employ discrete-event systems (DES) to model real systems
because it is suitable for describing dynamics and architectures of computer and
network systems~\cite{Cassandras2008}. We also utilize the fundamental techniques
from the supervisory control theory (SCT) of DES to compute solutions for
problems we introduce. The SCT is the theory that Ramadge and Wonham originally
proposed in~\cite{ramadge1987supervisory}. For a comprehensive account of the
SCT, the reader is referred to~\cite{wonham2018supervisory}, and also see
\cite{Wonham2018} for a historical overview of the theory.

One aspect of anonymity and secrecy that has been extensively studied in DES is
\textit{opacity}. This is a concept that intruders cannot identify secrets in
the system because of their partial observability. For an overview of opacity,
the reader is referred to~\cite{jacob2016overview}, and also
see~\cite{Lafortune2018257} for historical remarks on opacity. In case that
opacity is violated, several methods of enforcing opacity have been explored in
literature~\cite{Dubreil2008,Wu2015}. \cite{Dubreil2008}~investigates making
languages that the system generates opaque, namely intruders cannot determine
that the system has generated a secret language, by controlling the system with
the SCT. \cite{Wu2015}~introduces inserting observable events into output
languages from the system to make the language which intruders observe opaque.
Opacity is based on the side of intruders, assuming that they have full
knowledge of the target system's structure but have only partial observability
of the system's behavior. By contrast, our work in this paper stands on the side
of system administrators and focuses on the secret protection. We do not impose
assumptions on the intruders' knowledge and observability of the system. In
particular, intruders may be able to observe all events, in which case opacity
cannot hold. Instead, we study the problem of protecting the secrets as much as
possible, while balanced by the cost of such protections.

For protecting secrets, we consider that there exist some operations or events
which can be protected by administrators, e.g. connecting to a network or
logging into a computer. In this paper, we represent an event to which system
managers can apply a protection as a \textit{protectable event}, and other
events as \textit{unprotectable events}. We also consider that there are
multiple groups of protectable events, which have different levels of protection
implementation costs. In addition, we represent secret information to be
protected in the system as secret states. Secret information is a particular
piece of information which should be available only to permitted users, for
example, users' credit card numbers, or system privileges like \textit{root} in
Unix operating systems. System administrators decide which protectable events to
apply protections based on a \textit{protection policy} that specifies which
events to be protected at a given state. Our main objective is to solve the
problem of finding an effective protection policy such that all secrets are
protected with a predetermined number of protections, and the highest cost level
to implement these protections is minimum. To compute a solution for this
security problem, we convert the problem into a control problem and resort to
the SCT. Our previous work~\cite{Matsui2018} introduces a problem of secret
securing with at least one protection and minimum protection cost, which is a
special case of the problem considered in this paper.

The main contributions of this paper are fourfold. First, we formalize the
security problem with DES as secret protection with multiple protections and
minimum costs. Second, we present a necessary and sufficient condition under
which the problem is solvable. This condition characterizes the situation where
every string leading to the secret states in the system has at least a specified
number of protectable events. Third, we introduce the concept conversion from
security to control, and propose an algorithm to compute a solution for the
converted and the original problem.

The remaining of this paper is organized as follows. \cref{sec:formulation}
introduces a target system modeled by DES and formulates the problem of secret
securing with multiple protections and minimum costs. In \cref{sec:results}, we
first introduce a solvability condition such that the formulated original
problem is solvable, and convert the security problem to a control problem, and
then propose an algorithm to compute a solution for the converted problem.
\cref{sec:example} demonstrates our algorithm with an illustrating example.

\section{Problem Formulation}\label{sec:formulation}

In this section, we formulate ``Secret Securing with Multiple Protections and
Minimum Costs Problem''. Its objective is to find a policy to protect all secret
states with a prescribed number of protections and minimum protection cost.
Consider a task to protect all secrets in the system, and assume that
administrators want to use at least $m$ ($\geq 1$) protections. For this task, we
need to find a protection policy to force intruders before reaching secrets to
encounter $m$ protections. Meanwhile, the protection cost must be minimum. We
consider that all secrets are \textit{protected} with $m$ protections when every
string reaching secrets from the initial state has at least $m$ protectable
events.

We consider secret securing with minimum costs problem (SSMCP) in the framework
of discrete-event systems (DES) modeled as finite-state automata
\begin{equation}
  \label{eq:plant_sec}
  \mathbf{G} \coloneqq (Q, \Sigma, \delta, q_0)
\end{equation}
where $Q$ is the set of states, $\Sigma$ is the set of all events, $\delta: Q
\times \Sigma \to Q$ is the partial transition function, and $q_0 \in Q$ is the
initial state. We denote by $Q_s \subseteq Q$ the set of secret states in
$\mathbf{G}$. $\delta$ is extended to $\delta: Q \times \Sigma^\ast \to Q$ in
the standard manner~\cite{Cassandras2008}. $\delta(q, s)!$ denotes that string
$s$ from state $q$ is defined. $\Sigma$ is a disjoint union of the protectable
event set $\Sigma_p$ and the unprotectable event set $\Sigma_{up}$, namely
$\Sigma = \Sigma_p \disjoint \Sigma_{up}$. In addition, $\Sigma_p$ is
partitioned into $n$ disjoint subsets of protectable events $\Sigma_i$ where $i
\in \set{0, 1, \dots, n-1}$, namely $\Sigma_p = \bigdisjoint_{i=0}^{n-1}
\Sigma_i$. The index $i$ of $\Sigma_i$ indicates the level of protection cost
when the system administrator protects events in $\Sigma_i$. As the index $i$
increases, the protection cost becomes higher. We consider that the cost level
of each subset is not comparable with other subsets. In other words, the cost to
protect one event in $\Sigma_i$ is sufficiently higher than the cost to protect
all events in $\Sigma_{i-1}$. For example, implementing a biometric protection
is often more costly than setting up multiple password protections. We also
denote the union of the subsets of protectable events until index $k$ by
$\Sigma_p^k = \bigdisjoint_{i=0}^k \Sigma_i$.

In order to identify which transitions to protect, the system administrator
needs a protection policy which specifies protectable events at suitable states.
We define such a policy as a function $\mathcal{P}: Q \to \power(\Sigma_p)$
where $\power(\Sigma_p)$ is the power set of $\Sigma_p$. For example,
$\mathcal{P}(q) = \set{\sigma_i, \sigma_j}$ indicates that protectable events
$\sigma_i$ and $\sigma_j$ are protected at state $q$.

For clarity of presentation, we henceforth focus on the case $m = 2$. The case
$m \geq 3$ can be addressed in the same fashion (but with more complicated
notation). The case $m = 1$ has been solved in \cite{Matsui2018}, which is a
special case of the problem addressed in this paper. We first define the
following concept indicating that the secret states are protected with at least
two protections.

\begin{defn}[2-secure reachability]\label{defn:2-secure_reachability}
  Consider a plant $\mathbf{G}$ in \cref{eq:plant_sec}. The secret state $Q_s$
  is securely reachable with at least two protectable events (2-securely
  reachable) w.r.t. $\mathbf{G}$ and $\Sigma_p^k = \bigdisjoint_{i=0}^k
  \Sigma_i$ if the following condition holds:
  \begin{equation}\label{eq:defn:2-secure_reachability}
    [\forall s \in \Sigma^\ast] \delta(q_0, s)! \sand \delta(q_0, s) \in Q_s \implies s \in \Sigma^\ast\Sigma_p^k\Sigma^\ast\Sigma_p^k\Sigma^\ast
  \end{equation}
\end{defn}

Note that $s$ in \cref{eq:defn:2-secure_reachability} can contain two or more
protectable events in $\Sigma_p^k$, which means that intruders have to penetrate
at least two protections to reach secrets. When condition
\cref{eq:defn:2-secure_reachability} does not hold, there exists a string which
contains one or no protected event and reaches a secret state -- this is the
situation we try to avoid.

Next, we formulate the following security problem with
\cref{defn:2-secure_reachability}.

\begin{prob}[Secret Securing with Two Protections and Minimum Costs Problem, or $2$-SSMCP]\label{prob:2-ssmcp}
  Consider a plant $\mathbf{G}$ in \cref{eq:plant_sec}. Find a protection policy
  $\mathcal{P}: Q \to \power(\Sigma_p)$ s.t. $Q_s$ is 2-securely reachable
  w.r.t. $\Sigma_p^k = \bigdisjoint_{i=0}^k \Sigma_i$ and $k$ is the least
  index.
\end{prob}

Let us explain \cref{prob:2-ssmcp} with an illustrating example of a real
system.

\begin{exmp}\label{exmp:plant}
  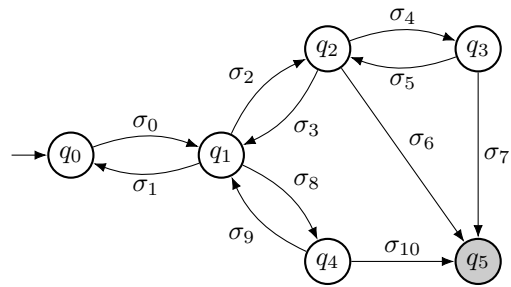
\begin{figure}[ht]
    \centering%
    \adjustbox{scale=1}{%
      \begin{tikzpicture}
  \node[state, initial] (0) {$q_0$};
  \node[state, right=of 0] (1) {$q_1$};
  \node[state, above right=of 1] (2) {$q_2$};
  \node[state, right=of 2] (3) {$q_3$};
  \node[state, below right=of 1] (4) {$q_4$};
  \node[state, right=of 4, fill=secret] (5) {$q_5$};

  \path
  (0) edge[bend left] node{$\sigma_0$} (1)
  (1) edge[bend left] node{$\sigma_1$} (0)
  (1) edge[bend left] node{$\sigma_2$} (2)
  (2) edge[bend left] node{$\sigma_3$} (1)
  (2) edge[bend left] node{$\sigma_4$} (3)
  (3) edge[bend left] node{$\sigma_5$} (2)
  (2) edge node{$\sigma_6$} (5)
  (3) edge node{$\sigma_7$} (5)
  (1) edge[bend left] node{$\sigma_8$} (4)
  (4) edge[bend left] node{$\sigma_9$} (1)
  (4) edge node{$\sigma_{10}$} (5)
  ;
\end{tikzpicture}%
    }
    \caption{Plant $\mathbf{G}$}
    \label{fig:exmp:plant}
  \end{figure}
  The plant $\mathbf{G}$ in \cref{fig:exmp:plant} represents a computer network
  composed of two different local-area networks (LAN) and two servers. Consider
  a situation where a user uses a laptop and wants to browse his or her secret
  file in the server via Wi-Fi. $q_0$ means ``the client is not connected'', and
  $q_1$ is a wireless router. Accordingly, $\sigma_0$ and $\sigma_1$ indicate
  connecting to and disconnecting from the router. $q_2$ and $q_4$ represent
  being in the respective LAN, that is, the client has been assigned an IP
  address. $\sigma_2$, $\sigma_3$, $\sigma_8$, and $\sigma_9$ are operations of
  connecting to and disconnecting from networks $q_2$ and $q_4$. $q_5$ is the
  server where the client's secret file is stored. $q_3$ is a bastion server
  different from $q_5$, which allows clients to access to $q_5$ through $q_3$,
  namely $\sigma_7$. Accordingly, $\sigma_4$ and $\sigma_5$ are logging on to
  and logging out from $q_3$. Users in LANs $q_2$ and $q_4$ can access to $q_5$
  directly. In order to protect the secret file against malicious access of
  intruders, the administrators must protect some suitable events in the plant,
  and they wish to do so with minimum cost.

  Let $\Sigma_p = \Sigma_0 \disjoint \Sigma_1 \disjoint \Sigma_2$ be the set of
  protectable events, where $\Sigma_0 = \set{\sigma_0}$, $\Sigma_1 =
  \set{\sigma_4, \sigma_6, \sigma_{10}}$, $\Sigma_2 = \set{\sigma_2, \sigma_8}$.
  Also let $\Sigma_{up} = \set{\sigma_1, \sigma_3, \sigma_5, \sigma_7,
    \sigma_9}$ be the set of unprotectable events, and $\Sigma = \Sigma_p
  \disjoint \Sigma_{up}$. The secret state $q_5$ is depicted as a shaded state
  in \cref{fig:exmp:plant}; thus $Q_s = \set{q_5}$. The $2$-SSMCP is the problem
  of finding a protection policy which specifies at least two protectable events
  in every path from $q_0$ to $q_5$ with minimum protection cost. In other
  words, every string reaching $q_5$ from $q_0$ must have at least two
  protectable events, and the index $k \in \set{0, 1, 2}$ of $\Sigma_p^k$ of
  these protectable events must be the smallest.
\end{exmp}

\section{Main Results}\label{sec:results}

In this section, we provide a necessary and sufficient condition for the
solvability of \cref{prob:2-ssmcp}, and compute a solution by resorting to the
SCT.

\subsection{Solvability of $2$-SSMCP}\label{subsec:solvability}

The following theorem provides a necessary and sufficient condition under which
there exists a solution of \cref{prob:2-ssmcp}.

\begin{thm}\label{thm:2-ssmcp_solvable}
  Consider a plant $\mathbf{G}$ in \cref{eq:plant_sec}. \cref{prob:2-ssmcp} is
  solvable w.r.t. $\mathbf{G}$ and $\Sigma_p^k = \bigdisjoint_{i=0}^k \Sigma_i$
  iff either
  \begin{gather}
    \text{$Q_s$ is 2-securely reachable w.r.t. $\mathbf{G}$ and
      $\Sigma_0$} \label{eq:thm:2-ssmcp_solvable:2} \\
    \intertext{or}
    \begin{aligned}
      &\left[ \text{$Q_s$ is 2-securely reachable w.r.t. $\mathbf{G}$ and $\Sigma_p^k$ \&} \right. \\
      &\left. \text{$Q_s$ is not 2-securely reachable w.r.t. $\mathbf{G}$ and $\Sigma_p^{k-1}$} \right]
    \end{aligned} \label{eq:thm:2-ssmcp_solvable:1}
  \end{gather}
  holds.
\end{thm}

Condition \cref{eq:thm:2-ssmcp_solvable:2} means that when $k = 0$, secret
states in $Q_s$ can be protected with at least two protections using protectable
events in $\Sigma_p^0 = \Sigma_0$. The meaning of
\cref{eq:thm:2-ssmcp_solvable:1} is that when $1 \leq k \leq n-1$, secret states
in $Q_s$ can be protected with at least two protections using protectable events
in $\Sigma_p^k$, and secrets can be protected with only one protection or cannot
be protected using protectable events only in $\Sigma_p^{k-1}$.

\begin{proof}
  ($\Rightarrow$) If \cref{eq:thm:2-ssmcp_solvable:2} is true, then $Q_s$ is
  2-securely reachable w.r.t. $\Sigma_p^0 = \Sigma_0$ (i.e. $k = 0$). The index
  $k$ of $\Sigma_p^k$ cannot be smaller than $0$, namely $k$ is minimum. In this
  case, there exists a protection policy $\mathcal{P}$ as a solution for
  \cref{prob:2-ssmcp} using protectable events only in $\Sigma_0$. Therefore, if
  \cref{eq:thm:2-ssmcp_solvable:2} holds, then \cref{prob:2-ssmcp} is solvable.
  Next, if \cref{eq:thm:2-ssmcp_solvable:1} is true, then $Q_s$ is 2-securely
  reachable w.r.t. $\Sigma_p^k$, and $k$ is the least index because $Q_s$ is not
  2-securely reachable w.r.t. $\Sigma_p^{k-1}$ and $\Sigma_p^{k-1} \subseteq
  \Sigma_p^k$. In this case, there exists a protection policy $\mathcal{P}$ as a
  solution for \cref{prob:2-ssmcp} using protectable events in $\Sigma_p^k$.
  Therefore, if \cref{eq:thm:2-ssmcp_solvable:1} holds, then \cref{prob:2-ssmcp}
  is solvable.

  ($\Leftarrow$) If \cref{prob:2-ssmcp} is solvable when $k = 0$, then $Q_s$ is
  2-securely reachable w.r.t. $\Sigma_0$. This is equivalent to
  \cref{eq:thm:2-ssmcp_solvable:2}. Thus if \cref{prob:2-ssmcp} is solvable when
  $k = 0$, then \cref{eq:thm:2-ssmcp_solvable:2} holds. Next, if
  \cref{prob:2-ssmcp} is solvable when $1 \leq k \leq n-1$, then $Q_s$ is
  2-securely reachable w.r.t. $\Sigma_p^k$, and since $k$ is minimum, $Q_s$ is
  not 2-securely reachable w.r.t. $\Sigma_p^{k-1}$ because of $\Sigma_p^{k-1}
  \subseteq \Sigma_p^k$. Therefore, if \cref{prob:2-ssmcp} is solvable when $1
  \leq k \leq n-1$, then \cref{eq:thm:2-ssmcp_solvable:1} holds.
\end{proof}

\subsection{Policy Computation}\label{subsec:computation}

In this subsection, we compute a protection policy when the solvability
condition of \cref{prob:2-ssmcp} in \cref{thm:2-ssmcp_solvable} holds. To
compute a protection policy, we convert the security problem
(\cref{prob:2-ssmcp}) to a control problem and resort to the SCT.
\begin{figure}[htp]
  \centering \adjustbox{scale=.9}{%
    \begin{tikzpicture}[node distance=1cm, concept/.style={thick, draw=black, minimum height=2em, minimum width=18ex}, trans/.style={->, shorten >=2pt, shorten <=2pt, >={Latex[length=8pt, width=8pt]}, line width=2pt}]
  \sffamily
  \node[concept, fill=black!20] (sec_prob) {Security Problem};
  \node[concept, below=17mm of sec_prob] (cont_prob) {Control Problem};
  \node[concept, right=46mm of cont_prob] (cont_policy) {Control Policy};
  \node[concept, above=17mm of cont_policy] (sec_policy) {Protection Policy};
  \node[concept, below=1 of cont_policy] (sup) {Supervisory Control};
  \node[right=23mm of cont_prob] (marker) {};

  \draw[-, thick] (sup.west) -| (marker.center);

  \path
  (sec_prob) edge[trans] (cont_prob)
  (cont_prob) edge[trans] (cont_policy)
  (cont_policy) edge[trans] (sec_policy)
  (sec_policy) edge[trans] (sec_prob)
  ;
\end{tikzpicture}%
  }
  \caption{Conversion overview}
  \label{fig:conversion}
\end{figure}
An overview of our concept conversions is shown in \cref{fig:conversion}. By the
conversion, protectable events $\Sigma_p$ and unprotectable events $\Sigma_{up}$
are converted to controllable events $\Sigma_c$ and uncontrollable events
$\Sigma_{uc}$ respectively. Accordingly, given a plant $\mathbf{G}$,
\cref{eq:plant_sec} becomes
\begin{equation}
  \label{eq:plant_cont}
  \mathbf{G} = (Q, \Sigma, \delta, q_0)
\end{equation}
where $\Sigma = \Sigma_c \disjoint \Sigma_{uc}$, and $\Sigma_c =
\bigdisjoint_{i=0}^{n-1} \Sigma_i$. Recall that $\Sigma_i$, $i \in \set{0, 1,
  \dots, n-1}$, denotes the partition of protectable events, indicating the cost
level with the index $i$. As $\Sigma_p^k$, we denote the union of the subsets of
controllable events until index $k$ by $\Sigma_c^k = \bigdisjoint_{i=0}^k
\Sigma_i$. Note that by the conversion, protection policy $\mathcal{P}$ is
converted to control policy $\mathcal{D}: Q \to \power(\Sigma_c)$ which is the
supervisor's decision of which controllable events to disable at any given
state. Letting $\mathbf{S} = (X, \Sigma, \xi, x_0)$ be a supervisor which is a
subautomaton of the plant $\mathbf{G}$, $\mathcal{D}$ is given by
\begin{equation}\label{eq:control_policy}
  \mathcal{D}(q) \coloneqq \begin{dcases}
    \set{\sigma \in \Sigma_c | \neg\xi(q, \sigma)! \sand \delta(q, \sigma)!} & \mathrm{if}\ q \in X \\
    \varnothing & \mathrm{if}\ q \in Q \setminus X
  \end{dcases}
\end{equation}

Based on the above conversion, the following definition and problem are
converted from \cref{defn:2-secure_reachability,prob:2-ssmcp}.

\begin{defn}[2-controllable reachability]\label{defn:2-controllable_reachability}
  Consider a plant $\mathbf{G}$ in \cref{eq:plant_cont}. The secret state set
  $Q_s$ is controllably reachable with at least two controllable events
  (2-controllably reachable) w.r.t. $\mathbf{G}$ and $\Sigma_c^k =
  \bigdisjoint_{i=0}^k \Sigma_i$ if the following condition holds:
  \begin{equation}\label{eq:defn:2-controllable_reachability}
    [\forall s \in \Sigma^\ast] \delta(q_0, s)! \sand \delta(q_0, s) \in Q_s \implies s \in \Sigma^\ast\Sigma_c^k\Sigma^\ast\Sigma_c^k\Sigma^\ast
  \end{equation}
\end{defn}

\begin{prob}[Reachability Control with Two Controllable Events and Minimum Costs
  Problem, or $2$-RCMCP]\label{prob:2-rcmcp}
  Consider a plant $\mathbf{G}$ in \cref{eq:plant_cont}. Find a control policy
  $\mathcal{D}$ s.t. $Q_s$ is 2-controllably reachable w.r.t. $\Sigma_c^k$ and
  $k$ is the least index.
\end{prob}

Next, the following is a necessary and sufficient condition under which
\cref{prob:2-rcmcp} is solvable.

\begin{prop}\label{prop:2-rcmcp_solvable}
  Consider a plant $\mathbf{G}$ in \cref{eq:plant_cont}. \cref{prob:2-rcmcp} is
  solvable iff either
  \begin{gather}
    \text{$Q_s$ is 2-controllably reachable w.r.t. $\mathbf{G}$ and $\Sigma_0$}
    \label{eq:prop:2-rcmcp_solvable:2} \\
    \intertext{or}
    \begin{aligned}
      &\left[ \text{$Q_s$ is 2-controllably reachable w.r.t. $\mathbf{G}$ and $\Sigma_c^k$ \&} \right. \\
      &\left. \text{$Q_s$ is not 2-controllably reachable w.r.t. $\mathbf{G}$
          and $\Sigma_c^{k-1}$} \right]
    \end{aligned} \label{eq:prop:2-rcmcp_solvable:1}
  \end{gather}
  holds.
\end{prop}

\begin{proof}
  The proof follows from \cref{thm:2-ssmcp_solvable} and the conversion.
\end{proof}

\begin{prob}[Reachability Control with One Controllable Event and Minimum Cost
  Problem, or $1$-RCMCP]\label{prob:1-rcmcp}
  Consider a plant $\mathbf{G}$ in \cref{eq:plant_cont}. Find a control policy
  $\mathcal{D}$ s.t. $Q_s$ is controllably reachable with at least one
  controllable event w.r.t. $\Sigma_c^k$ and $k$ is the least index.
\end{prob}

The following relation holds between \cref{prob:1-rcmcp} and \cref{prob:2-rcmcp}.

\begin{prop}\label{prop:1-rcmcp_2-rcmcp}
  \cref{prob:1-rcmcp} is solvable if \cref{prob:2-rcmcp} is solvable.
\end{prop}

\begin{proof}
  \cref{prob:1-rcmcp} is solvable iff either
  \begin{gather}
    [\forall s \in (\Sigma\setminus\Sigma_0)^\ast]\delta(q_0, s)
    \not\in Q_s \label{eq:proof:prop:1-rcmcp_2-rcmcp:1} \\
    \intertext{or}
    \begin{aligned}
      &\big[ [\forall s \in (\Sigma\setminus\Sigma_c^k)^\ast]\delta(q_0, s) \not\in Q_s \land \\
      &\qquad [\exists s \in (\Sigma\setminus\Sigma_c^{k-1})^\ast]\delta(q_0, s)
      \in Q_s \big]
    \end{aligned} \label{eq:proof:prop:1-rcmcp_2-rcmcp:2}
  \end{gather}
  holds (cf. \cite{Matsui2018}). \cref{eq:proof:prop:1-rcmcp_2-rcmcp:1} is
  equivalent to
  \begin{equation}\label{eq:proof:prop:1-rcmcp_2-rcmcp:3}
    [\forall s \in \Sigma^\ast] \delta(q_0, s)! \sand \delta(q_0, s) \in Q_s \implies s \in \Sigma^\ast\Sigma_0\Sigma^\ast
  \end{equation}
  and \cref{eq:proof:prop:1-rcmcp_2-rcmcp:2} is equivalent to
  \begin{equation}\label{eq:proof:prop:1-rcmcp_2-rcmcp:4}
    \begin{gathered}
      [\forall s \in \Sigma^\ast] \delta(q_0, s)! \sand \delta(q_0, s) \in Q_s \implies s \in \Sigma^\ast\Sigma_c^k\Sigma^\ast \\
      \sand \\
      [\exists s \in \Sigma^\ast] \delta(q_0, s)! \sand \delta(q_0, s) \in Q_s \sand s \not\in \Sigma^\ast\Sigma_c^{k-1}\Sigma^\ast
    \end{gathered}
  \end{equation}
  Moreover from \cref{defn:2-controllable_reachability},
  \cref{eq:prop:2-rcmcp_solvable:2} is equivalent to
  \begin{equation}\label{eq:proof:prop:1-rcmcp_2-rcmcp:5}
    [\forall s \in \Sigma^\ast] \delta(q_0, s)! \sand \delta(q_0, s) \in Q_s \implies s \in \Sigma^\ast\Sigma_0\Sigma^\ast\Sigma_0\Sigma^\ast
  \end{equation}
  and \cref{eq:prop:2-rcmcp_solvable:1} is equivalent to
  \begin{equation}\label{eq:proof:prop:1-rcmcp_2-rcmcp:6}
    \begin{gathered}
      [\forall s \in \Sigma^\ast] \delta(q_0, s)! \sand \delta(q_0, s) \in Q_s
      \implies s \in \Sigma^\ast\Sigma_c^k\Sigma^\ast\Sigma_c^k\Sigma^\ast \\
      \sand \\
      [\exists s \in \Sigma^\ast] \delta(q_0, s)! \sand \delta(q_0, s) \in Q_s \sand s \not\in \Sigma^\ast\Sigma_c^{k-1}\Sigma^\ast\Sigma_c^{k-1}\Sigma^\ast
    \end{gathered}
  \end{equation}
  From \cref{prop:2-rcmcp_solvable}, if \cref{prob:2-rcmcp} is solvable when $k
  = 0$, then \cref{eq:proof:prop:1-rcmcp_2-rcmcp:5} is true. Thus from
  $\Sigma^\ast\Sigma_0\Sigma^\ast\Sigma_0\Sigma^\ast \subseteq
  \Sigma^\ast\Sigma_0\Sigma^\ast$, \cref{eq:proof:prop:1-rcmcp_2-rcmcp:3} is
  also true. Thus from \cref{eq:proof:prop:1-rcmcp_2-rcmcp:1}, when $k = 0$,
  \cref{prob:1-rcmcp} is solvable if \cref{prob:2-rcmcp} is solvable. In the
  same way, when $1 \leq k \leq n-1$, \cref{prob:1-rcmcp} is solvable if
  \cref{prob:2-rcmcp} is solvable from \cref{eq:proof:prop:1-rcmcp_2-rcmcp:2},
  \cref{eq:proof:prop:1-rcmcp_2-rcmcp:4}, and
  \cref{eq:proof:prop:1-rcmcp_2-rcmcp:6}.
\end{proof}

To compute a control policy which specifies at least two controllable events in
every string reaching secret states from the initial state, we propose
\cref{alg:rcmc2}. This algorithm computes two supervisors $\mathbf{S}_0$ and
$\mathbf{S}_1$ for $\mathbf{G}$ in \cref{eq:plant_cont}. Each supervisor
provides a different control policy such that every string reaching secret
states has at least one controllable event with minimum cost. To compute the
first supervisor $\mathbf{S}_0$, we design the control specification
$\mathbf{G}_K$ by removing the secret states in $Q_s$ and the transitions to and
from removed secret states:
\begin{equation}
  \label{eq:spec}
  \mathbf{G}_K = (Q \setminus Q_s, \Sigma, \delta_K, q_0)
\end{equation}
where $\delta_K = \delta \setminus \set{(q, \sigma, q^\prime) | q, q^\prime \in
  Q_s, \sigma \in \Sigma}$.

Note that in real systems, secret states are still reachable. It is not suitable
to disable events to protect secrets because it can inhibit users' normal
behavior.

\begin{algorithm}[htp]
  \caption{RCMC2}
  \label{alg:rcmc2}
  \begin{algorithmic}[1]
    \Require{$\mathbf{G}$ in \cref{eq:plant_cont}, $\mathbf{G}_K$ in
      \cref{eq:spec}}%
    \Ensure{Supervisor automata $\mathbf{S}_0$ and $\mathbf{S}_1$}%
    \State{Compute $\mathbf{S}_0$, $k_0$ by RCMC1 with inputs $\mathbf{G}$,
      $\mathbf{G}_K$}%
    \If{$\mathbf{S}_0$ is nonempty}%
    \State{Derive $\mathcal{D}_0$ from $\mathbf{S}_0$ by
      \cref{eq:control_policy}}%
    \State{Form $\mathbf{G}_1 = (Q, \Sigma^1, \delta^1, q_0)$ as in
      \cref{eq:relabeled_plant}}%
    \State{$\delta_K^1 = \delta^1 \setminus \set{(q, \sigma, q^\prime) | q,
        q^\prime \in Q_s, \sigma \in \Sigma^1}$}%
    \State{$\mathbf{G}_{K_1} = (Q \setminus Q_s, \Sigma^1, \delta_K^1, q_0)$}%
    \State{Compute $\mathbf{S}_1$, $k_1$ by RCMC1 with inputs $\mathbf{G}_1$,
      $\mathbf{G}_{K_1}$}%
    \State \Return{$\mathbf{S}_0$, $\mathbf{S}_1$}%
    \EndIf%
    \State \Return{Empty supervisors}%
    \State%
    \Function{RCMC1}{$\mathbf{G}$, $\mathbf{G}_K$}%
    \State{$K = L(\mathbf{G}_K)$}%
    \For{$k = 0, 1, \dots, n - 1$}%
    \State{$\Sigma_c^k = \displaystyle \bigdisjoint^k_{i = 0} \Sigma_i$}%
    \State{Compute a supervisor $\mathbf{S}$ s.t. $L(\mathbf{S}) = \supc(K)$
      w.r.t. $\Sigma_c^k$}%
    \If{$\mathbf{S}$ is nonempty}%
    \State \Return{$\mathbf{S}$, $k$}%
    \EndIf%
    \EndFor%
    \State \Return{Empty supervisor, null}%
    \EndFunction%
  \end{algorithmic}
\end{algorithm}

\begin{exmp}\label{exmp:spec}
  \begin{figure}[ht]
    \centering%
    \adjustbox{scale=1}{%
      \begin{tikzpicture}
  \node[state, initial] (0) {$q_0$};
  \node[state, right=of 0] (1) {$q_1$};
  \node[state, above right=of 1] (2) {$q_2$};
  \node[state, right=of 2] (3) {$q_3$};
  \node[state, below right=of 1] (4) {$q_4$};

  \path
  (0) edge[bend left] node{$\sigma_0$} (1)
  (1) edge[bend left] node{$\sigma_1$} (0)
  (1) edge[bend left] node{$\sigma_2$} (2)
  (2) edge[bend left] node{$\sigma_3$} (1)
  (2) edge[bend left] node{$\sigma_4$} (3)
  (3) edge[bend left] node{$\sigma_5$} (2)
  (1) edge[bend left] node{$\sigma_8$} (4)
  (4) edge[bend left] node{$\sigma_9$} (1)
  ;
\end{tikzpicture}%
    }
    \caption{Specification $\mathbf{G}_K$}
    \label{fig:exmp:spec}
  \end{figure}
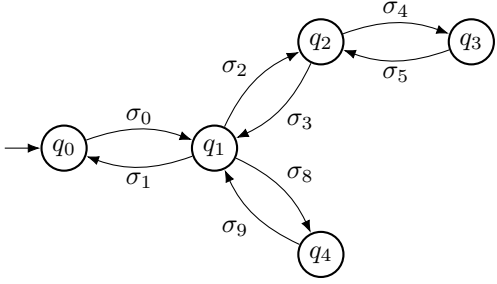
  The specification automaton for the plant in \cref{exmp:plant} is shown in
  \cref{fig:exmp:spec}. From \cref{eq:spec}, secret state $Q_s = \set{q_5}$ and
  transitions $(q_2, \sigma_6, q_5)$, $(q_3, \sigma_7, q_5)$, and $(q_4,
  \sigma_{10}, q_5)$ are removed from the plant $\mathbf{G}$ in
  \cref{fig:exmp:plant}.
\end{exmp}

To proceed, we need several standard concepts of the SCT. Consider a plant
$\mathbf{G}$ in \cref{eq:plant_cont}. Let $K = L(\mathbf{G}_K) \subseteq
L(\mathbf{G})$ be a specification language. From \cite{wonham2018supervisory},
$K$ is controllable if $\overline{K}\Sigma_{uc} \cap L(\mathbf{G}) \subseteq
\overline{K}$ where $\overline{K}$ is the prefix closure of $K$. The family
$\mathcal{C}(K)$ is the set of all controllable sublanguages of $K$, and denoted
by $\mathcal{C}(K) \coloneqq \set{K^\prime \subseteq K |
  \overline{K^\prime}\Sigma_{uc} \cap L(\mathbf{G}) \subseteq
  \overline{K^\prime}}$. The supremal controllable sublanguage of $K$ is given
by $\supc(K) \coloneqq \bigcup\set{K^\prime | K^\prime \in \mathcal{C}(K)}$.
$\supc(K)$ w.r.t. $\mathbf{G}$ and $\Sigma_c^k$ means $\supc(K) =
\bigcup\set{K^{\prime\prime} | K^{\prime\prime} \in \mathcal{C}(K)}$ where
$\mathcal{C}(K) = \set{K^\prime \subseteq K | \overline{K^\prime} (\Sigma
  \setminus \Sigma_c^k) \cap L(\mathbf{G}) \subseteq \overline{K^\prime}}$.

\begin{lem}\label{lem:supc}
  (cf. \cite{wonham2018supervisory}) Let $\mathbf{G} = (Q, \Sigma_{uc} \disjoint
  \Sigma_c, \delta, q_0)$ be a plant and $K \subseteq L(\mathbf{G})$ be a
  specification language. The following holds:
  \begin{equation}\label{eq:lem:supc}
    \supc(K) = \varnothing \iff [\exists s \in \Sigma_{uc}^\ast] s \in L(\mathbf{G}) \setminus K
  \end{equation}
\end{lem}

From \cref{lem:supc} and the construction of $\mathbf{G}_K$ in \cref{eq:spec},
letting $K = L(\mathbf{G}_K)$, the supervisor $\mathbf{S}_0 = \supc(K)$ w.r.t.
$\mathbf{G}$ in \cref{eq:plant_cont} and $\Sigma_c^k$ is nonempty if and only if
every string which contains events in $\Sigma_c^k$ in $\mathbf{G}$ and reaches
secret states has at least one controllable event. In other words, $\supc(K)
\neq \varnothing$ w.r.t. $\mathbf{G}$ and $\Sigma_c^k$ if and only if $[\forall
s \in (\Sigma \setminus \Sigma_c^k)^\ast]\delta(q_0, s) \not\in Q_s$.
Accordingly, the RCMC1 function in \cref{alg:rcmc2} returns a supervisor which
specifies controllable events such that every string reaching secret states from
the initial state has at least one controllable event. The index $k$ which RCMC1
returns is minimum because the index in RCMC1 starts from $0$ and is incremented
by $1$ at each iteration.

Let $\mathcal{D}_0$ be a control policy derived from the first supervisor
$\mathbf{S}_0$ as in \cref{eq:control_policy}. To compute the second supervisor
$\mathbf{S}_1$, we relabel the transitions specified by $\mathcal{D}_0$ to
distinguish the disabled transitions and other non-disabled transitions.
Relabeled controllable transitions are treated as uncontrollable. Accordingly, a
new plant $\mathbf{G}_1$ is defined as follows:
\begin{align}
  \mathbf{G}_1 &= (Q, \Sigma^1, \delta^1, q_0) \label{eq:relabeled_plant} \\
  \Sigma^1 &= \Sigma_{uc_1} \disjoint\,(\Sigma_c \setminus \set{\sigma \in \Sigma | (q, \sigma, q^\prime) \in \delta_{\mathcal{D}_0}}) \\
  \Sigma_{uc_1} &= \Sigma_{uc} \disjoint \set{\sigma \in \Sigma | (q, \sigma^\prime, q^\prime) \in \delta_{\mathcal{D}_0}^\prime} \\
  \delta_{\mathcal{D}_0} &= \set{(q, \sigma, q^\prime) | q, q^\prime \in Q, \sigma \in \mathcal{D}_0(q)} \label{eq:transition_in_policy} \\
  \delta_{\mathcal{D}_0}^\prime &= \set{(q, \sigma^\prime, q^\prime) | q, q^\prime \in Q, \sigma \in \mathcal{D}_0(q)} \label{eq:relabeled_transition} \\
  \delta^1 &= (\delta \setminus \delta_{\mathcal{D}_0}) \disjoint \delta_{\mathcal{D}_0}^\prime \label{eq:new_transition}
\end{align}
Note that $\Sigma_{uc_1}$ is the subset of uncontrollable events in
$\mathbf{G}_1$. We call the sequence from \cref{eq:relabeled_plant} to
\cref{eq:new_transition} that defines $\mathbf{G}_1$ ``relabeling''.

\begin{exmp}\label{exmp:relabel}
  Consider the plant $\mathbf{G}$ in \cref{exmp:plant}. The control policy
  $\mathcal{D}_0$ derived from the first supervisor for \cref{exmp:plant} is as
  follows:
  \begin{equation}\label{eq:exmp:relabel:1}
    \begin{aligned}
      \mathcal{D}_0(q_0) &= \set{\sigma_0} \\
      \mathcal{D}_0(q_1) &= \mathcal{D}_0(q_2) = \mathcal{D}_0(q_3) = \varnothing \\
      \mathcal{D}_0(q_4) &= \mathcal{D}_0(q_4) = \mathcal{D}_0(q_5) =
      \varnothing
    \end{aligned}
  \end{equation}
  \begin{figure}[htp]
    \centering \adjustbox{scale=1}{%
      \begin{tikzpicture}
  \node[state, initial] (0) {$q_0$};
  \node[state, right=of 0] (1) {$q_1$};
  \node[state, above right=of 1] (2) {$q_2$};
  \node[state, right=of 2] (3) {$q_3$};
  \node[state, below right=of 1] (4) {$q_4$};
  \node[state, right=of 4, fill=secret] (5) {$q_5$};

  \path
  (0) edge[bend left] node[label={[shift={(-8mm,-5pt)}]0:\faLock}] {$\sigma_0$} (1)
  (1) edge[bend left] node{$\sigma_1$} (0)
  (1) edge[bend left] node{$\sigma_2$} (2)
  (2) edge[bend left] node{$\sigma_3$} (1)
  (2) edge[bend left] node{$\sigma_4$} (3)
  (3) edge[bend left] node{$\sigma_5$} (2)
  (2) edge node{$\sigma_6$} (5)
  (3) edge node{$\sigma_7$} (5)
  (1) edge[bend left] node{$\sigma_8$} (4)
  (4) edge[bend left] node{$\sigma_9$} (1)
  (4) edge node{$\sigma_{10}$} (5)
  ;
\end{tikzpicture}%
    }
    \caption{The plant $\mathbf{G}$ with the protection policy $\mathcal{P}_0$}
    \label{fig:exmp:relabel}
  \end{figure}
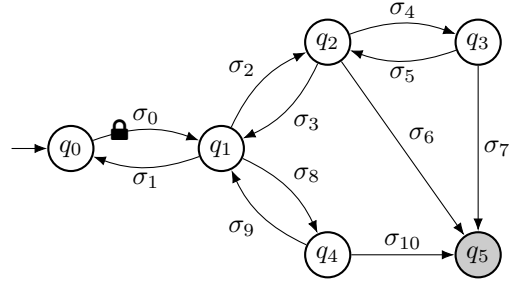
  The protection policy $\mathcal{P}_0$ derived from $\mathcal{D}_0$ by the
  conversion is shown in \cref{fig:exmp:relabel}. ``\faLock'' means that the
  event is protected. Before we compute the second supervisor to obtain a
  solution for \cref{prob:2-rcmcp}, we relabel the disabled transitions
  $\mathcal{D}_0$ specifies as follows:
  \begin{equation}
    \delta_{\mathcal{D}_0}^\prime = \set{(q_0, \sigma_0^\prime, q_1)}
  \end{equation}
  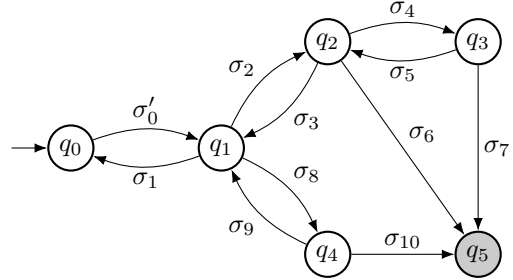
\begin{figure}[htp]
    \centering \adjustbox{scale=1}{%
      \begin{tikzpicture}
  \node[state, initial] (0) {$q_0$};
  \node[state, right=of 0] (1) {$q_1$};
  \node[state, above right=of 1] (2) {$q_2$};
  \node[state, right=of 2] (3) {$q_3$};
  \node[state, below right=of 1] (4) {$q_4$};
  \node[state, right=of 4, fill=secret] (5) {$q_5$};

  \path
  (0) edge[bend left] node{$\sigma_0^\prime$} (1)
  (1) edge[bend left] node{$\sigma_1$} (0)
  (1) edge[bend left] node{$\sigma_2$} (2)
  (2) edge[bend left] node{$\sigma_3$} (1)
  (2) edge[bend left] node{$\sigma_4$} (3)
  (3) edge[bend left] node{$\sigma_5$} (2)
  (2) edge node{$\sigma_6$} (5)
  (3) edge node{$\sigma_7$} (5)
  (1) edge[bend left] node{$\sigma_8$} (4)
  (4) edge[bend left] node{$\sigma_9$} (1)
  (4) edge node{$\sigma_{10}$} (5)
  ;
\end{tikzpicture}%
    }
    \caption{Plant $\mathbf{G}_1$}
    \label{fig:exmp:relabeled_plant}
  \end{figure}
  Based on this relabeling, a new plant $\mathbf{G}_1 = (Q, \Sigma^1, \delta^1,
  q_0)$ derived from $\mathbf{G}$ in \cref{fig:exmp:plant} is shown in
  \cref{fig:exmp:relabeled_plant}.

  Finally, letting $\delta^1_K = \delta^1 \setminus \set{(q, \sigma, q^\prime) |
    q, q^\prime \in Q_s, \sigma \in \Sigma^1}$, we design the specification
  automaton $\mathbf{G}_{K_1} = (Q \setminus Q_s, \Sigma^1, \delta^1_K, q_0)$
  for the relabeled plant $\mathbf{G}_1$ to compute the second supervisor
  $\mathbf{S}_1$. For \cref{exmp:plant}, $\mathbf{G}_{K_1}$ is depicted in
  \cref{fig:spec_relabeled}.
  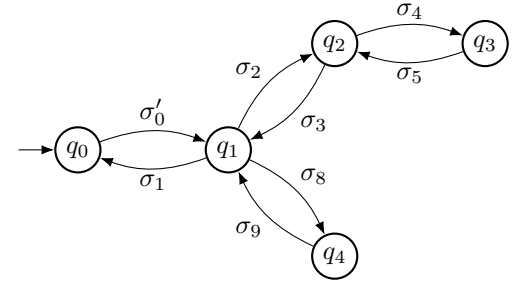
\begin{figure}[htp]
    \centering%
    \adjustbox{scale=1}{%
      \begin{tikzpicture}
  \node[state, initial] (0) {$q_0$};
  \node[state, right=of 0] (1) {$q_1$};
  \node[state, above right=of 1] (2) {$q_2$};
  \node[state, right=of 2] (3) {$q_3$};
  \node[state, below right=of 1] (4) {$q_4$};

  \path
  (0) edge[bend left] node{$\sigma_0^\prime$} (1)
  (1) edge[bend left] node{$\sigma_1$} (0)
  (1) edge[bend left] node{$\sigma_2$} (2)
  (2) edge[bend left] node{$\sigma_3$} (1)
  (2) edge[bend left] node{$\sigma_4$} (3)
  (3) edge[bend left] node{$\sigma_5$} (2)
  (1) edge[bend left] node{$\sigma_8$} (4)
  (4) edge[bend left] node{$\sigma_9$} (1)
  ;
\end{tikzpicture}%
    }
    \caption{Specification $\mathbf{G}_{K_1}$}
    \label{fig:spec_relabeled}
  \end{figure}
\end{exmp}

\cref{alg:rcmc2} returns either empty or nonempty supervisor automata
$\mathbf{S}_0$ and $\mathbf{S}_1$. If \cref{alg:rcmc2} returns two nonempty
supervisors, there exists a control policy $\mathcal{D}_0$ by the supervisor
$\mathbf{S}_0$ such that $L(\mathbf{S}_0) = \supc(L(\mathbf{G}_{K}))$ and
$\mathcal{D}_1$ by $\mathbf{S}_1$ such that $L(\mathbf{S}_1) =
\supc(L(\mathbf{G}_{K_1}))$. From $\mathcal{D}_0$ and $\mathcal{D}_1$, a
solution for \cref{prob:2-rcmcp}, namely $\mathcal{D}: Q \to \power(\Sigma_c)$,
is given by
\begin{equation}\label{eq:merge_cont_policy}
  \mathcal{D}(q) \coloneqq \mathcal{D}_0(q) \disjoint \mathcal{D}_1(q)
\end{equation}
In other words, \cref{eq:merge_cont_policy} means merging $\mathcal{D}_0$ and
$\mathcal{D}_1$. Each control policy specifies controllable events such that
every string reaching secrets has at least one controllable event. Therefore,
$\mathcal{D}$ in \cref{eq:merge_cont_policy} specifies at least two controllable
events in every string reaching secret states from the initial state.

Index $k_1$ in \cref{alg:rcmc2} line 7 is equal to or larger than $k_0$ in line
1, namely $k_0 \leq k_1$. This is because $k_0$ is the least index such that
there exists a control policy $\mathcal{D}_0$ to make secret states unreachable
in $\mathbf{G}$. Moreover, letting $\Sigma_c^k$ be the subset of controllable
events that $\mathcal{D}$ in \cref{eq:merge_cont_policy} specifies to disable,
index $k$ is minimum because $k_1$ are minimum and $k = k_1$. Thus, the
condition in \cref{prop:2-rcmcp_solvable} is satisfied with $k = k$ in
\cref{eq:prop:2-rcmcp_solvable:1,eq:prop:2-rcmcp_solvable:2}.

\begin{prop}\label{prop:relabel_2-rcmcp}
  \cref{alg:rcmc2} returns nonempty supervisors iff \cref{prob:2-rcmcp} is
  solvable.
\end{prop}

\begin{proof}
  ($\Leftarrow$) From \cref{prop:1-rcmcp_2-rcmcp}, if \cref{prob:2-rcmcp} is
  solvable, then a supervisor $\mathbf{S}_0$ on line 2 of \cref{alg:rcmc2} is
  nonempty. Furthermore from \cref{prop:2-rcmcp_solvable}, if
  \cref{prob:2-rcmcp} is solvable, then every string $s \in \Sigma^\ast$ s.t.
  $\delta(q_0, s) \in Q_s$ contains two or more controllable events. Thus from
  the definition of relabeling in
  \cref{eq:relabeled_plant}--\cref{eq:new_transition}, every string $s \in
  \Sigma^\ast$ s.t. $\delta^1(q_0, s) \in Q_s$ has at least one controllable
  event. Therefore, there exists an index $k_1$ (where $k_0 \leq k_1 \leq n-1$)
  of $\Sigma_c^{k_1}$ s.t. $\supc(L(\mathbf{G}_{K_1})) \neq \varnothing$ w.r.t.
  $\mathbf{G}_1$ and $\Sigma_c^{k_1}$, and RCMC1 returns a nonempty supervisor,
  namely $\mathbf{S}_1$ on line 7 of \cref{alg:rcmc2} is nonempty. It follows
  from \cref{alg:rcmc2} line 8 that the nonempty supervisors $\mathbf{S}_0$ and
  $\mathbf{S}_1$ are returned, and index $k_1$ is the least.

  ($\Rightarrow$) When \cref{alg:rcmc2} returns nonempty supervisors,
  $\mathbf{S}_0$ and $\mathbf{S}_1$ in \cref{alg:rcmc2} are nonempty. Thus from
  the relabeling and RCMC1, if $k_0 = k_1 = 0$, each of $\mathcal{D}_0$ and
  $\mathcal{D}_1$ specifies controllable events belonging to $\Sigma_0$.
  Therefore, letting $k = k_1 = 0$, condition \cref{eq:prop:2-rcmcp_solvable:2}
  is true. Hence from \cref{prop:2-rcmcp_solvable}, \cref{prob:2-rcmcp} is
  solvable if \cref{alg:rcmc2} returns nonempty supervisors and $k_0 = k_1 =
  0$. Likewise, when
  \begin{equation}\label{eq:proof:relabel_2-rcmcp}
    0 \leq k_0 \leq k_1 \leq n-1
  \end{equation}
  in \cref{alg:rcmc2}, letting $k = k_1$, \cref{eq:prop:2-rcmcp_solvable:1} is
  true because $\Sigma_c^{k_0} \subseteq \Sigma_c^{k_1}$. Hence from
  \cref{prop:2-rcmcp_solvable}, \cref{prob:2-rcmcp} is solvable if
  \cref{alg:rcmc2} returns nonempty supervisors and
  \cref{eq:proof:relabel_2-rcmcp} is true.
\end{proof}

From \cref{eq:merge_cont_policy}, a solution for \cref{prob:2-ssmcp}, namely
$\mathcal{P}: Q \to \power(\Sigma_p)$, is given by
\begin{equation}\label{eq:merge_prot_policy}
  \mathcal{P}(q) \coloneqq \mathcal{P}_0(q) \disjoint \mathcal{P}_1(q)
\end{equation}
where $\mathcal{P}_0$ and $\mathcal{P}_1$ are derived from $\mathcal{D}_0$ and
$\mathcal{D}_1$ respectively by inverse conversion. The least index is $k =
k_1$.

Finally, we state our main result.

\begin{thm}\label{thm:rcmc2}
  Consider a plant $\mathbf{G}$ in \cref{eq:plant_sec}. If \cref{prob:2-ssmcp}
  is solvable, then the protection policy $\mathcal{P}$ in
  \cref{eq:merge_prot_policy} is a solution for \cref{prob:2-ssmcp}.
\end{thm}

\begin{proof}
  Suppose that \cref{prob:2-ssmcp} is solvable. Then \cref{prob:2-rcmcp} is
  solvable by conversion of protectable events to controllable events. Then by
  \cref{prop:relabel_2-rcmcp}, \cref{alg:rcmc2} returns nonempty supervisors
  $\mathbf{S}_0$ and $\mathbf{S}_1$ such that $L(\mathbf{S}_0) =
  \supc(L(\mathbf{G}_K))$ and $L(\mathbf{S}_1) = \supc(L(\mathbf{G}_{K_1}))$.
  Based on $\mathbf{S}_0$ and $\mathbf{S}_1$, control policies $\mathcal{D}_0$
  and $\mathcal{D}_1$ can be defined as in \cref{eq:control_policy}
  respectively. Thus a merged control policy $\mathcal{D}$ can be defined as in
  \cref{eq:merge_cont_policy} from $\mathcal{D}_0$ and $\mathcal{D}_1$. From the
  relabeling, $\mathcal{D}_0$ and $\mathcal{D}_1$ specify different transitions
  to disable. Also it follows from $\mathbf{G}_K$ and $\mathbf{G}_{K_1}$ that
  $Q_s$ is controllably reachable under each of $\mathcal{D}_0$ and
  $\mathcal{D}_1$. Therefore, under control policy $\mathcal{D}$, $Q_s$ is
  2-controllably reachable. Moreover, letting $k = k_1$ (where $k_1$ is from
  line 7 of \cref{alg:rcmc2}) and $\Sigma_c^k$ be the subset of controllable
  events that $\mathcal{D}$ specifies, index $k$ of $\Sigma_c^k$ is minimum
  because $k = k_1$ and $k_1$ is minimum. Hence the control policy $\mathcal{D}$
  derived from $\mathcal{D}_0$ and $\mathcal{D}_1$ computed by \cref{alg:rcmc2}
  is a solution for \cref{prob:2-rcmcp}. Consequently from the conversion, the
  protection policy $\mathcal{P}$ defined in \cref{eq:merge_prot_policy} is a
  solution for \cref{prob:2-ssmcp}.
\end{proof}

\subsection{Securing with Multiple Protections}\label{subsec:multiple}

When administrators need to protect secrets with strictly more than two
protections, $2$-SSMCP (\cref{prob:2-ssmcp}) is extended to $m$ ($\geq 3$)
protections ($m$-SSMCP). To compute a solution for $m$-SSMCP, we iterate the
relabeling procedure and the function RCMC1 in \cref{alg:rcmc2} until secrets
are protected by $m$ protections. Letting $j$ be the execution count of RCMC1,
the relabeling procedure \cref{eq:relabeled_plant}--\cref{eq:new_transition} is
also extended for $m$-SSMCP as follows:
\begin{align}
  \mathbf{G}_{j+1} &= (Q, \Sigma^{j+1}, \delta^{j+1}, q_0) \label{eq:multiple_plant} \\
  \Sigma^{j+1} &= \Sigma_{uc_{j+1}} \disjoint\,(\Sigma_c \setminus \set{\sigma \in \Sigma | (q, \sigma, q^\prime) \in \delta_{\mathcal{D}_j}}) \\
  \Sigma_{uc_{j+1}} &= \Sigma_{uc} \disjoint \set{\sigma \in \Sigma | (q, \sigma^\prime, q^\prime) \in \delta_{\mathcal{D}_j}^\prime} \\
  \delta_{\mathcal{D}_j} &= \set{(q, \sigma, q^\prime) | q, q^\prime \in Q, \sigma \in \mathcal{D}_j(q)} \\
  \delta_{\mathcal{D}_j}^\prime &= \set{(q, \sigma^\prime, q^\prime) | q, q^\prime \in Q, \sigma \in \mathcal{D}_j(q)} \\
  \delta^{j+1} &= (\delta \setminus \delta_{\mathcal{D}_j}) \disjoint \delta_{\mathcal{D}_j}^\prime
\end{align}
Based on this extension, to compute a protection policy such that every string
reaching secrets has $m$ protectable events, we propose
\cref{alg:m-rcmc} as an extension of \cref{alg:rcmc2}.
\begin{algorithm}[htp]
  \caption{RCMC$m$}
  \label{alg:m-rcmc}
  \begin{algorithmic}[1]
    \Require{$\mathbf{G}$ in \cref{eq:plant_cont}, $\mathbf{G}_K$ in
      \cref{eq:spec}, $m$}%
    \Ensure{Supervisor automata $\mathbf{S}_0$, $\mathbf{S}_1$, \dots,
      $\mathbf{S}_{m-1}$}%
    \State{$\mathbf{G}_0 = \mathbf{G}, \mathbf{G}_{K_0} = \mathbf{G}_K$}%
    \For{$j = 0, 1, \dots, m-1$}%
    \State{Compute $\mathbf{S}_j$, $k_j$ by RCMC1 in \cref{alg:rcmc2} with
      inputs $\mathbf{G}_j$, $\mathbf{G}_{K_j}$}%
    \If{$\mathbf{S}_j$ is nonempty}%
    \State{Derive $\mathcal{D}_j$ from $\mathbf{S}_j$ by
      \cref{eq:control_policy}}%
    \State{Form $\mathbf{G}_{j+1} = (Q, \Sigma^{j+1}, \delta^{j+1}, q_0)$ from
      $\mathbf{G}_j$ and $\mathcal{D}_j$ as in \cref{eq:multiple_plant}}%
    \State{$\delta_K^{j+1} = \delta^{j+1} \setminus \set{(q, \sigma, q^\prime) | q,
      q^\prime \in Q_s, \sigma \in \Sigma^{j+1}}$}%
    \State{$\mathbf{G}_{K_{j+1}} = (Q \setminus Q_s, \Sigma^{j+1},
      \delta_K^{j+1}, q_0)$}%
    \Else%
    \State\Return{Empty supervisors}%
    \EndIf%
    \EndFor%
    \State\Return{$\mathbf{S}_0$, $\mathbf{S}_1$, \dots, $\mathbf{S}_{m-1}$}
  \end{algorithmic}
\end{algorithm}

\section{Illustrating Example}\label{sec:example}

\subsection{Two Protections}\label{subsec:example_2-ssmcp}

Let us take \cref{exmp:plant} again to demonstrate our developed solution for
$2$-SSMCP.

Consider the plant $\mathbf{G}$ in \cref{exmp:plant}. We first convert $2$-SSMCP
(\cref{prob:2-ssmcp}) to $2$-RCMCP (\cref{prob:2-rcmcp}) by converting
protectable events to controllable events. In \cref{alg:rcmc2} line 1, from
\cref{exmp:relabel}, $\mathbf{S}_0$ is nonempty. In line 3, $\mathcal{D}_0$ is
in \cref{eq:exmp:relabel:1}. In line 4, the new plant $\mathbf{G}_1$ derived
from $\mathbf{G}$ and $\mathcal{D}_0$ by the relabeling is in
\cref{fig:exmp:relabeled_plant}. In line 6, the specification automaton
$\mathbf{G}_{K_1}$ for $\mathbf{G}_1$ is shown in \cref{fig:spec_relabeled}. Call
function RCMC1 at line 7. In line 13, let $K = L(\mathbf{G}_{K_1})$. In line 14,
initially $k = 0$:
\begin{align*}
  \Sigma_c^0 &= \set{\sigma_0} \\
  \Sigma \setminus \Sigma_c^0 &= \Sigma_{uc1} \disjoint \Sigma_1 \disjoint \Sigma_2 \\
  \shortintertext{Then}
  \mathcal{C}(K) &= \set{\varnothing} \\
  \shortintertext{Hence}
  \supc(K) &= \varnothing \\
  \intertext{Thus increment $k$ by $1$, i.e. $k = 1$:}
  \Sigma_c^1 &= \set{\sigma_0, \sigma_4, \sigma_6, \sigma_{10}} \\
  \Sigma \setminus \Sigma_c^1 &= \Sigma_{uc1} \disjoint \Sigma_2 \\
  \shortintertext{Then}
  \mathcal{C}(K) &= \set{\varnothing, (\sigma_0^\prime(\sigma_2.\sigma_3)^\ast(\sigma_8.\sigma_9)^\ast\sigma_1)^\ast} \\
  \shortintertext{Hence}
  \supc(K) &= (\sigma_0^\prime(\sigma_2.\sigma_3)^\ast(\sigma_8.\sigma_9)^\ast\sigma_1)^\ast
\end{align*}
So in line 18, function RCMC1 returns $\mathbf{S}_1$ with
$L(\mathbf{S}_1) = \supc(K)$, and $k = 1$. Then, \cref{alg:rcmc2} returns
nonempty supervisors $\mathbf{S}_0$ and $\mathbf{S}_1$. According to
$\mathbf{S}_1$, a control policy $\mathcal{D}_1$ for $\mathbf{G}_1$ is as
follows:
\begin{equation}\label{eq:second_policy}
  \begin{aligned}
    \mathcal{D}_1(q_2) &= \set{\sigma_4, \sigma_6} \\
    \mathcal{D}_1(q_4) &= \set{\sigma_{10}} \\
    \mathcal{D}_1(q_0) &= \mathcal{D}_1(q_1) = \mathcal{D}_1(q_3) =
    \mathcal{D}_1(q_5) = \varnothing
  \end{aligned}
\end{equation}
Therefore, the solution for \cref{prob:2-rcmcp} for this example is the
following control policy $\mathcal{D}$ derived from
\cref{eq:exmp:relabel:1,eq:second_policy} by \cref{eq:merge_cont_policy}:
\begin{equation}\label{eq:exmp_cont_policy}
  \begin{aligned}
    \mathcal{D}(q_0) &= \set{\sigma_0} \\
    \mathcal{D}(q_2) &= \set{\sigma_4, \sigma_6} \\
    \mathcal{D}(q_4) &= \set{\sigma_{10}} \\
    \mathcal{D}(q_1) &= \mathcal{D}(q_3) = \mathcal{D}(q_5) = \varnothing
  \end{aligned}
\end{equation}

Finally, the solution for \cref{prob:2-ssmcp} (i.e. the protection policy
$\mathcal{P}$) for this example is derived from \cref{eq:exmp_cont_policy} by
the reverse conversion:
\begin{equation}\label{eq:exmp_sec_policy}
  \begin{aligned}
    \mathcal{P}(q_0) &= \set{\sigma_0} \\
    \mathcal{P}(q_2) &= \set{\sigma_4, \sigma_6} \\
    \mathcal{P}(q_4) &= \set{\sigma_{10}} \\
    \mathcal{P}(q_1) &= \mathcal{P}(q_3) = \mathcal{P}(q_5) = \varnothing
  \end{aligned}
\end{equation}

\begin{figure}[htp]
  \centering \adjustbox{scale=1}{%
    \begin{tikzpicture}
  \node[state, initial] (0) {$q_0$};
  \node[state, right=of 0] (1) {$q_1$};
  \node[state, above right=of 1] (2) {$q_2$};
  \node[state, right=of 2] (3) {$q_3$};
  \node[state, below right=of 1] (4) {$q_4$};
  \node[state, right=of 4, fill=secret] (5) {$q_5$};

  \path
  (0) edge[bend left] node[label={[shift={(-8mm,-5pt)}]0:\faLock}] {$\sigma_0$} (1)
  (1) edge[bend left] node{$\sigma_1$} (0)
  (1) edge[bend left] node{$\sigma_2$} (2)
  (2) edge[bend left] node{$\sigma_3$} (1)
  (2) edge[bend left] node[label={[shift={(-8mm,-5pt)}]0:\faLock}]{$\sigma_4$} (3)
  (3) edge[bend left] node{$\sigma_5$} (2)
  (2) edge node[label={[shift={(0,1mm)}]180:\faLock}]{$\sigma_6$} (5)
  (3) edge node{$\sigma_7$} (5)
  (1) edge[bend left] node{$\sigma_8$} (4)
  (4) edge[bend left] node{$\sigma_9$} (1)
  (4) edge node[label={[shift={(-9mm,-4pt)}]0:\faLock}]{$\sigma_{10}$} (5)
  ;
\end{tikzpicture}%
  }
  \caption{The plant $\mathbf{G}$ with the protection policy $\mathcal{P}$}
  \label{fig:exmp_solution}
\end{figure}
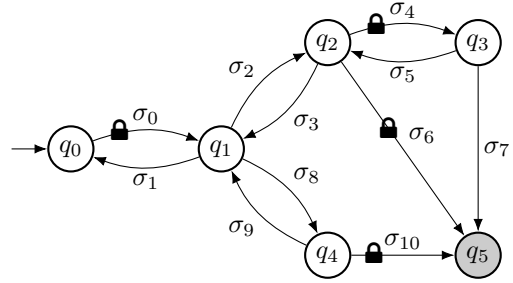

\cref{fig:exmp_solution} illustrates the plant $\mathbf{G}$ with the protection
policy $\mathcal{P}$. From $\mathcal{P}$, the secret $q_5$ is 2-securely
reachable because all strings from the initial state $q_0$ reaching the secret
state have at least two protectable events. For example, in real systems,
protecting $\sigma_0$ can be implemented by setting up a Wi-Fi password for the
wireless router. Additionally, protections for $\sigma_4$, $\sigma_6$ and
$\sigma_{10}$ can be implemented by configuration of user authentication in each
of the servers $q_3$ and $q_5$.

\subsection{Three Protections}\label{subsec:example_3-ssmcp}

In this subsection, we demonstrate our solution for $m$-SSMCP for the case $m =
3$. Consider again the plant $\mathbf{G}$ in \cref{exmp:plant}, and assume that
the secret in $\mathbf{G}$ must be protected by at least three protections,
namely $m = 3$ of $m$-SSMCP. The transitions in $\mathbf{G}$ which
$\mathcal{D}_0$ and $\mathcal{D}_1$ specify to disable are relabeled as follows:
\begin{equation*}
  \delta_{\mathcal{D}_1}^\prime = \set{(q_0, \sigma_0^\prime, q_1), (q_2, \sigma_4^\prime, q_3), (q_2, \sigma_6^\prime, q_5), (q_4, \sigma_{10}^\prime, q_5)}
\end{equation*}
\begin{figure}[htp]
  \centering \adjustbox{scale=1}{%
    \begin{tikzpicture}
  \node[state, initial] (0) {$q_0$};
  \node[state, right=of 0] (1) {$q_1$};
  \node[state, above right=of 1] (2) {$q_2$};
  \node[state, right=of 2] (3) {$q_3$};
  \node[state, below right=of 1] (4) {$q_4$};
  \node[state, right=of 4, fill=secret] (5) {$q_5$};

  \path
  (0) edge[bend left] node{$\sigma_0^\prime$} (1)
  (1) edge[bend left] node{$\sigma_1$} (0)
  (1) edge[bend left] node{$\sigma_2$} (2)
  (2) edge[bend left] node{$\sigma_3$} (1)
  (2) edge[bend left] node{$\sigma_4^\prime$} (3)
  (3) edge[bend left] node{$\sigma_5$} (2)
  (2) edge node{$\sigma_6^\prime$} (5)
  (3) edge node{$\sigma_7$} (5)
  (1) edge[bend left] node{$\sigma_8$} (4)
  (4) edge[bend left] node{$\sigma_9$} (1)
  (4) edge node{$\sigma_{10}^\prime$} (5)
  ;
\end{tikzpicture}%
  }
  \caption{Plant $\mathbf{G}_2$}
  \label{fig:exmp:3-ssmcp:relabeled}
\end{figure}
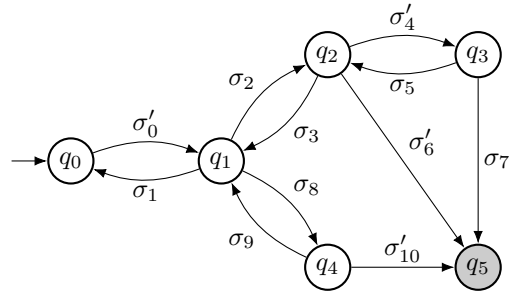
Next, the new plant $\mathbf{G}_2$ is depiceted in
\cref{fig:exmp:3-ssmcp:relabeled}. After computing the specification automaton
$\mathbf{G}_{K_2}$ for $\mathbf{G}_2$ in the same way as $\mathbf{G}_{K_1}$, let
$K = L(\mathbf{G}_{K_2})$ and call function RCMC1 in line 3 of
\cref{alg:m-rcmc}. When $k = 0, 1$, $\supc(K) = \varnothing$. When $k = 2$,
\begin{equation*}
  \supc(K) = (\sigma_0^\prime.\sigma_1)^\ast
\end{equation*}
So RCMC1 returns $\mathbf{S}_2$ with $L(\mathbf{S}_2) = \supc(K)$. Thus
\cref{alg:m-rcmc} returns nonempty supervisors $\mathbf{S}_0$, $\mathbf{S}_1$,
$\mathbf{S}_2$. According to $\mathbf{S}_2$, the following is a control policy
$\mathcal{D}_2$ for $\mathbf{G}_2$:
\begin{align*}
  \mathcal{D}_2(q_1) &= \set{\sigma_2, \sigma_8} \\
  \mathcal{D}_2(q_0) &= \mathcal{D}_2(q_2) = \mathcal{D}_2(q_3) = \mathcal{D}_2(q_4) = \mathcal{D}_2(q_5) = \varnothing
\end{align*}
Therefore, the merged control policy $\mathcal{D}$ for this example is as
follows, derived from $\mathcal{D}_0$, $\mathcal{D}_1$ and $\mathcal{D}_2$:
\begin{align*}
  \mathcal{D}(q_0) &= \set{\sigma_0} \\
  \mathcal{D}(q_1) &= \set{\sigma_2, \sigma_8} \\
  \mathcal{D}(q_2) &= \set{\sigma_4, \sigma_6} \\
  \mathcal{D}(q_4) &= \set{\sigma_{10}} \\
  \mathcal{D}(q_3) &= \mathcal{D}(q_5) = \varnothing
\end{align*}

Finally, the solution for $m$-SSMCP (i.e. the protection policy $\mathcal{P}$)
for this example is derived from $\mathcal{D}$ by the reverse conversion:
\begin{align*}
  \mathcal{P}(q_0) &= \set{\sigma_0} \\
  \mathcal{P}(q_1) &= \set{\sigma_2, \sigma_8} \\
  \mathcal{P}(q_2) &= \set{\sigma_4, \sigma_6} \\
  \mathcal{P}(q_4) &= \set{\sigma_{10}} \\
  \mathcal{P}(q_3) &= \mathcal{P}(q_5) = \varnothing
\end{align*}

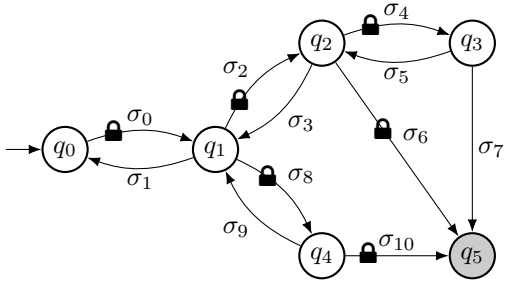
\begin{figure}[htp]
  \centering \adjustbox{scale=1}{%
    \begin{tikzpicture}
  \node[state, initial] (0) {$q_0$};
  \node[state, right=of 0] (1) {$q_1$};
  \node[state, above right=of 1] (2) {$q_2$};
  \node[state, right=of 2] (3) {$q_3$};
  \node[state, below right=of 1] (4) {$q_4$};
  \node[state, right=of 4, fill=secret] (5) {$q_5$};

  \path
  (0) edge[bend left] node[label={[shift={(-8mm,-5pt)}]0:\faLock}] {$\sigma_0$} (1)
  (1) edge[bend left] node{$\sigma_1$} (0)
  (1) edge[bend left] node[label={[shift={(-4mm,-4mm)}]0:\faLock}]{$\sigma_2$} (2)
  (2) edge[bend left] node{$\sigma_3$} (1)
  (2) edge[bend left] node[label={[shift={(-8mm,-5pt)}]0:\faLock}]{$\sigma_4$} (3)
  (3) edge[bend left] node{$\sigma_5$} (2)
  (2) edge node[label={[shift={(0,1mm)}]180:\faLock}]{$\sigma_6$} (5)
  (3) edge node{$\sigma_7$} (5)
  (1) edge[bend left] node[label={[shift={(0pt,0pt)}]180:\faLock}]{$\sigma_8$} (4)
  (4) edge[bend left] node{$\sigma_9$} (1)
  (4) edge node[label={[shift={(-9mm,-4pt)}]0:\faLock}]{$\sigma_{10}$} (5)
  ;
\end{tikzpicture}%
  }
  \caption{The plant $\mathbf{G}$ with the protection policy $\mathcal{P}$}
  \label{fig:exmp:3-ssmcp}
\end{figure}
\cref{fig:exmp:3-ssmcp} illustrates the plant $\mathbf{G}$ with the protection
policy $\mathcal{P}$. From $\mathcal{P}$, every string reaching the secret state
$q_5$ from the initial state $q_0$ has at least three protectable events, that
is, the secret is protected with at least three protections. For example,
protecting $\sigma_2$ and $\sigma_8$ can be implemented with Authentication VLAN
(IEEE 802.1X) \cite{rfc:3580} to prevent users from accessing the prohibited
network in real systems. However, installing and configuring VLAN for the system
is generally much more difficult for network administrators than setting
connection passwords of the wireless router, or than creating accounts for users
in the servers.

\section{Conclusions}\label{sec:conclusions}

We have introduced the problem of protecting secret states in the system with at
least $m$ ($\geq 1$) protections and minimum protection costs ($m$-SSMCP). This
problem has been formulated as finding a protection policy such that every
string reaching secret states from the initial state has at least $m$
protectable events, and the protection cost is minimum. We have presented a
solution algorithm for $m$-SSMCP which computes $m$ supervisors. Finally, we
have demonstrated our solution with a network example.

In future work, we aim to investigate a situation where secrets have different
importance and administrators are concerned with the balance between protection
cost and secret importance.

\bibliographystyle{IEEEtran}
\bibliography{IEEEabrv,reference}

\begin{thebibliography}{10}
\providecommand{\url}[1]{#1}
\csname url@rmstyle\endcsname
\providecommand{\newblock}{\relax}
\providecommand{\bibinfo}[2]{#2}
\providecommand\BIBentrySTDinterwordspacing{\spaceskip=0pt\relax}
\providecommand\BIBentryALTinterwordstretchfactor{4}
\providecommand\BIBentryALTinterwordspacing{\spaceskip=\fontdimen2\font plus
\BIBentryALTinterwordstretchfactor\fontdimen3\font minus
  \fontdimen4\font\relax}
\providecommand\BIBforeignlanguage[2]{{%
\expandafter\ifx\csname l@#1\endcsname\relax
\typeout{** WARNING: IEEEtran.bst: No hyphenation pattern has been}%
\typeout{** loaded for the language `#1'. Using the pattern for}%
\typeout{** the default language instead.}%
\else
\language=\csname l@#1\endcsname
\fi
#2}}

\bibitem{brooks2018cybersecurity}
C.~J. Brooks, C.~Grow, P.~Craig, and D.~Short, \emph{Cybersecurity
  Essentials}.\hskip 1em plus 0.5em minus 0.4em\relax John Wiley \& Sons, 2018.

\bibitem{Cassandras2008}
C.~Cassandras and S.~Lafortune, \emph{{Introduction to Discrete Event
  Systems}}.\hskip 1em plus 0.5em minus 0.4em\relax Boston, MA: Springer US,
  2008.

\bibitem{ramadge1987supervisory}
P.~J. Ramadge and W.~M. Wonham, ``Supervisory control of a class of discrete
  event processes,'' \emph{SIAM Journal on Control and Optimization}, vol.~25,
  no.~1, pp. 206--230, 1987.

\bibitem{wonham2018supervisory}
W.~M. Wonham and K.~Cai, \emph{Supervisory Control of Discrete-Event
  Systems}.\hskip 1em plus 0.5em minus 0.4em\relax Springer International
  Publishing, 2018.

\bibitem{Wonham2018}
W.~Wonham, K.~Cai, and K.~Rudie, ``{Supervisory control of discrete-event
  systems: A brief history},'' \emph{Annual Reviews in Control}, vol.~45, pp.
  250--256, 2018.

\bibitem{jacob2016overview}
R.~Jacob, J.~J. Lesage, and J.~M. Faure, ``{Overview of Discrete Event Systems
  Opacity: models, validation and quantification},'' \emph{Annual Reviews in
  Control}, vol.~28, no.~7, pp. 174--181, 2015.

\bibitem{Lafortune2018257}
S.~Lafortune, F.~Lin, and C.~N. Hadjicostis, ``On the history of diagnosability
  and opacity in discrete event systems,'' \emph{Annual Reviews in Control},
  vol.~45, pp. 257--266, 2018.

\bibitem{Dubreil2008}
J.~Dubreil, P.~Darondeau, and H.~Marchand, ``{Opacity enforcing control
  synthesis},'' in \emph{Proceedings of the 9th International Workshop on
  Discrete Event Systems}, 2008, pp. 28--35.

\bibitem{Wu2015}
Y.~C. Wu and S.~Lafortune, ``{Synthesis of opacity-enforcing insertion
  functions that can be publicly known},'' in \emph{Proceedings of the 54th
  IEEE Conference on Decision and Control}, 2015, pp. 3506--3513.

\bibitem{Matsui2018}
S.~Matsui and K.~Cai, ``Secret securing with minimum cost,'' in
  \emph{Proceedings of the 61st Japan Joint Automatic Control Conference},
  2018, pp. 1017--1024.

\bibitem{rfc:3580}
P.~Congdon, B.~Aboba, A.~Smith, G.~Zorn, and J.~Roese, ``{I}{E}{E}{E} 802.1{X}
  remote authentication dial in user service ({R}{A}{D}{I}{U}{S}) usage
  guidelines,'' RFC 3580, Sept. 2003.

\end{thebibliography}

\end{document}